\documentclass[%
 reprint,
 showkeys,
 amsmath,amssymb,
 aps,
]{revtex4-2}

\usepackage{graphicx}
\usepackage{color}
\usepackage[normalem]{ulem}
\usepackage{dcolumn}
\usepackage{bm}
\usepackage[usenames,dvipsnames]{xcolor}

\begin{document}

\preprint{APS/123-QED}

\title{Transfer reactions of exotic nuclei including core deformations: \textsuperscript{11}Be and \textsuperscript{17}C }

\author{P. Punta}
 \email{ppunta@us.es}
 \affiliation{%
 Departamento de FAMN, Facultad de Física, Universidad de Sevilla, Apartado 1065, E-41080 Sevilla, Spain
}%
\author{J. A. Lay}
 \email{lay@us.es}
\author{A. M. Moro}
 \email{moro@us.es}
\affiliation{%
 Departamento de FAMN, Facultad de Física, Universidad de Sevilla, Apartado 1065, E-41080 Sevilla, Spain
}%
\affiliation{Instituto Interuniversitario Carlos I de F\'isica Te\'orica y Computacional (iC1), Apdo.~1065, E-41080 Sevilla, Spain}

\date{\today}

\begin{abstract}
\begin{description}
\item[Background] Reactions with halo nuclei from deformed regions exhibit important deviations from the inert core+valence picture. Structure and reaction formalisms have recently been extended or adapted to explore the possibility of exciting the underlying core.
\item[Purpose] We will study up to what extent transfer reactions involving halo nuclei $^{11}$Be and $^{17}$C can be reproduced with two different models that have previously shown a good success reproducing the role of the core in light halo nuclei.

\item[Methods] We focus on the structure of $^{11}$Be and $^{17}$C with two core+valence models: Nilsson and a semimicroscopic particle-rotor model using Antisymmetrized Molecular Dinamic calculations of the cores. These models are later used to study $^{16}\text{C}(d,p)^{17}\text{C}$ and $^{11}\text{Be}(p,d)^{10}\text{Be}$ transfer reactions within the Adiabatic Distorted Wave Approximation. Results are compared with three different experimental data sets.

\item[Results] A good reproduction of both the structure and transfer reactions of $^{11}$Be and $^{17}$C is found. The Nilsson model provides an overall better agreement for the spectrum and reactions involving $^{17}$C while the semi-microscopic model is more adequate for $^{11}$Be, as expected,  since the $^{17}$C core is closer to an ideal rotor. 
\item[Conclusions] Both models show promising results for the study of transfer reactions with halo nuclei. We expect that including microscopic information in the Nilsson model, following the spirit of the semi-microscopic model, can provide a useful, yet simple framework for studying newly discovered halo nuclei.
\end{description}
\end{abstract}

\maketitle

\section{\label{sec:level1}Introduction}

The study of exotic nuclei is one of the main topics in
current nuclear physics research. 
They are nuclei far from the stability line, 
with a rather different ratio of protons to neutrons
from that of stable nuclei. 
Because of that, they usually
exhibit very different properties from those of stable nuclei.
A particularly interesting case is that of halo nuclei. 
These are weakly bound systems composed of
one or two hihgly delocalized valence particle(s)
and a relatively compact core.
As a schematic picture, 
the valence particles form a \textit{halo} of matter around the core.

Weakly bound nuclei are conveniently described within
few-body models,
in which deformations of the fragments are usually ignored.
However, \textit{core deformations} are known to affect significantly both
the structure and the dynamics of these systems{~\cite{Mor12,Mor12a,Del13,Lay16}}.
Therefore, deformation needs to be included
in structure and reaction models,
for a meaningful and reliable description of reactions including these nuclei.

In previous works, the effect of core deformations in nuclear reactions has been included within the particle-rotor model, which is based on a weak-coupling limit. This model may be inaccurate for well-deformed nuclei, for which approaches based on the strong-coupling limit, such as the well-known Nilsson model,  might provide a more suitable framework.

In this work, we present an exploratory study of the deformed weakly-bound nuclei $^{17}$C and $^{11}$Be within the Nilsson model. A two-body model is considered: a neutron moving in a deformed potential generated by the core. Although this is not the first application of the Nilsson model to these nuclei \cite{Ham07,Ham07_2}, a novelty of our work is the use of a pseudo-state method to compute the bound and unbound states of the system. In this method, the energies of these states and their associated wavefunctions are obtained diagonalizing the internal Hamiltonian
in a basis of square-integrable functions, for which we employ the transformed harmonic oscillator functions (THO) which has been successfully applied to the discretization
of the continuum of weakly bound nuclei
for its application to  breakup and transfer direct reactions 
both for two-body and three-body systems \cite{Mor12a,Casal16}.

For comparison purposes, we also present calculations for the same nuclei based on the so-called PAMD model, in which the Hamiltonian is constructed using the transition densities of the corresponding cores calculated in the Antisymmetrized Molecular Dynamics (AMD) formalism. The PAMD model was introduced in \cite{Lay14} and applied to $^{11}\text{Be}$ and $^{19}\text{C}$ with a promising reproduction of the structure of both nuclei. Results for $^{17}$C can be found in \cite{Punta,Pereira}.

To assess the quality and reliability of the developed models, the calculated wavefunctions are applied to transfer reactions involving these nuclei. 
Two transfer reactions have been studied by
implementing the results of the structure calculations, 
$^{16}\text{C}(d,p)^{17}\text{C}$ and $^{11}\text{Be}(p,d)^{10}\text{Be}$.
The adiabatic distorted wave approximation (ADWA) \cite{ADWA} has been used for this purpose.
The results of theoretical calculations are compared with
the experimental data recently measured in GANIL and RCNP.

The Nilsson scheme has been previously applied to reactions involving light exotic nuclei \cite{Macc17,Macc18,Macc18erratum} but only to extract the required spectroscopic factors, which are then combined with single-particle wavefunctions calculated in a spherical potential. 
However, to the best of our knowledge, this is the first time that the effect of the deformation is considered also on the radial form factor of the transfer reaction for exotic nuclei and, not only, for the calculation of the corresponding spectroscopic factor.  As already shown in stable nuclei~\cite{Bro76}, there is a non-negligible effect on the radial extension of the form factor due to the deformation. One can only expect this effect to increase considerably in the case of halo nuclei.

The paper is organized as follows. Sec.~II. is about the structure formalism,
it focuses on the description of the novel approach based on the Nilsson model
and also briefly explains the PAMD model.
Sec. III shows the application of the ADWA approach to one neutron transfer reactions.
The results of the application of the two models to the $^{17}$C
and $^{11}$Be nuclei can be seen in Sec.~IV.
In Sec.~V we study the reactions
$^{16}\text{C}(d,p)^{17}\text{C}$ and $^{11}\text{Be}(p,d)^{10}\text{Be}$.
Finally, we discuss the main results in Sec.~VI.

\section{\label{sec:level1}Structure Formalism}
We consider a composite nucleus, described as a two-body system,
comprising a weakly-bound nucleon coupled to a core.
The Hamiltonian of the system can be written as 
\begin{equation}\label{eq:H}
{\cal H}=T(\vec r)+V_{\ell s}(r)(\vec \ell\cdot\vec s)+V_{vc}(\vec r,\xi)+h_{core}(\xi).
\end{equation}
where $T(\vec r)$ is the kinetic energy operator for the relative
motion between the valence and the core,
$h_{core}(\xi)$ is the Hamiltonian of the core,
and $V_{vc}(\vec r,\xi)$ is the effective valence-core interaction.
A spin-obit term with the usual radial dependence $V_{\ell s}(r)$
is added to this valence-core interaction.
$\xi$ denotes the core degrees of freedom,
so the dependence of $V_{vc}(\vec r,\xi)$
on it accounts for core-excitation effects.

In this work, two different models have been considered,
which can be regarded as opposite limits of the coupling strength, namely, strong- and weak-coupling.
For the strong-coupling, the Nilsson model, as formulated in Ref.~\cite{Hamamoto05}, was used. For the weak coupling case, we employ the semi-microscopic particle-plus-AMD (PAMD) model proposed in Ref.~\cite{Lay14}. 
This second model obtains the coupling potential $V_{vc}(\vec r,\xi)$ 
convoluting an effective NN interaction 
with microscopic transition densities of the core nucleus  calculated with 
Antysymmetrized Molecular Dynamics (AMD)  \cite{Kan95a}.

The eigenfunctions of the Hamiltonian, for a given energy $\varepsilon$,
are characterized by the parity $\pi$ 
and the total angular momentum $\vec{J}$,
resulting from the coupling of the angular momentum $\vec{j}$
of the valence particle to the core angular momentum $\vec{I}$.
These functions can be generically expressed as
\begin{equation}\label{eq:LabWF}
\Psi_{\varepsilon M}^{J^\pi}(\vec{r},\xi)=\sum_{\alpha}
R^{J^\pi}_{\varepsilon\alpha}(r)\Phi_{\alpha J}^{M}(\hat{r},\xi),
\end{equation}
where $\Phi_{\alpha J}^M(\hat{r},\xi)$ refers to the eigenstates of $J^2$ and $J_z$ 
resulting for the coupling of $\vec j$ to $\vec I$,
\begin{equation}\label{eq:LabAng}
\Phi_{\alpha J}^M(\hat{r},\xi)\equiv\left[  
{\cal Y}_{\ell s}^j(\hat{r}) \otimes \phi_{I}(\xi) \right]_{JM}.
\end{equation}
Here, $\vec{\ell}$ is the orbital
angular momentum of the valence particle relative to the core,
which couples to the spin of the valence particle $\vec s$ to give the
particle total angular momentum $\vec{j}$.
The label $\alpha$ denotes the set of quantum numbers $\{\ell,s,j,I\}$.
${\cal Y}_{\ell s}^{jm}(\hat{r})$ denotes the wavefunction resulting
from coupling the spin of the valence particle with the corresponding spherical harmonic.

\subsection{\label{sec:level2}Nilsson model}
A key aspect of the Nilsson model is that, instead of considering
the relative motion valence-core in the space fixed laboratory frame ($\vec r$),
it considers the intrinsic frame ($\vec{r'}$), which rotates jointly with the core.
For this frame, if we assume that the core has a permanent deformation,
we can assume that the potential $V_{vc}$ depends on $\vec{r'}$ with the same geometry
and does not depend explicitly on $\xi$.

In the original Nilsson model, the valence-core interaction
is assumed to be an anisotropic harmonic oscillator potential.
However, in this work, a more realistic Woods-Saxon potential is used
and a permanent axially symmetric quadrupole deformation is applied.
Following  \cite{Hamamoto05}, 
we obtain to first order in the deformation parameter $\beta$:
\begin{equation}\label{eq:NilssonPot}
 V_{vc}^{\text{Nilsson}}(r,\theta')=
 V_c(r)-\beta r\frac{dV_c(r)}{dr}Y_{20}(\theta'),
\end{equation}
where $\theta'$ is the angle with respect to the symmetry axis of the core
and $r'$ coincides with $r$, the relative distance between core and valence.
Note that deformations with $\beta>0$ and $\beta<0$  correspond, respectively, to prolate and oblate shapes.
Expression (\ref{eq:NilssonPot}) can be in principle applied to deform any central potential $V_c(r)$,
and, in particular, if an isotropic harmonic oscillator potential is used,
the original Nilsson model term for the anisotropic oscillator is recovered.

\begin{figure}
\includegraphics[width=0.9\linewidth]{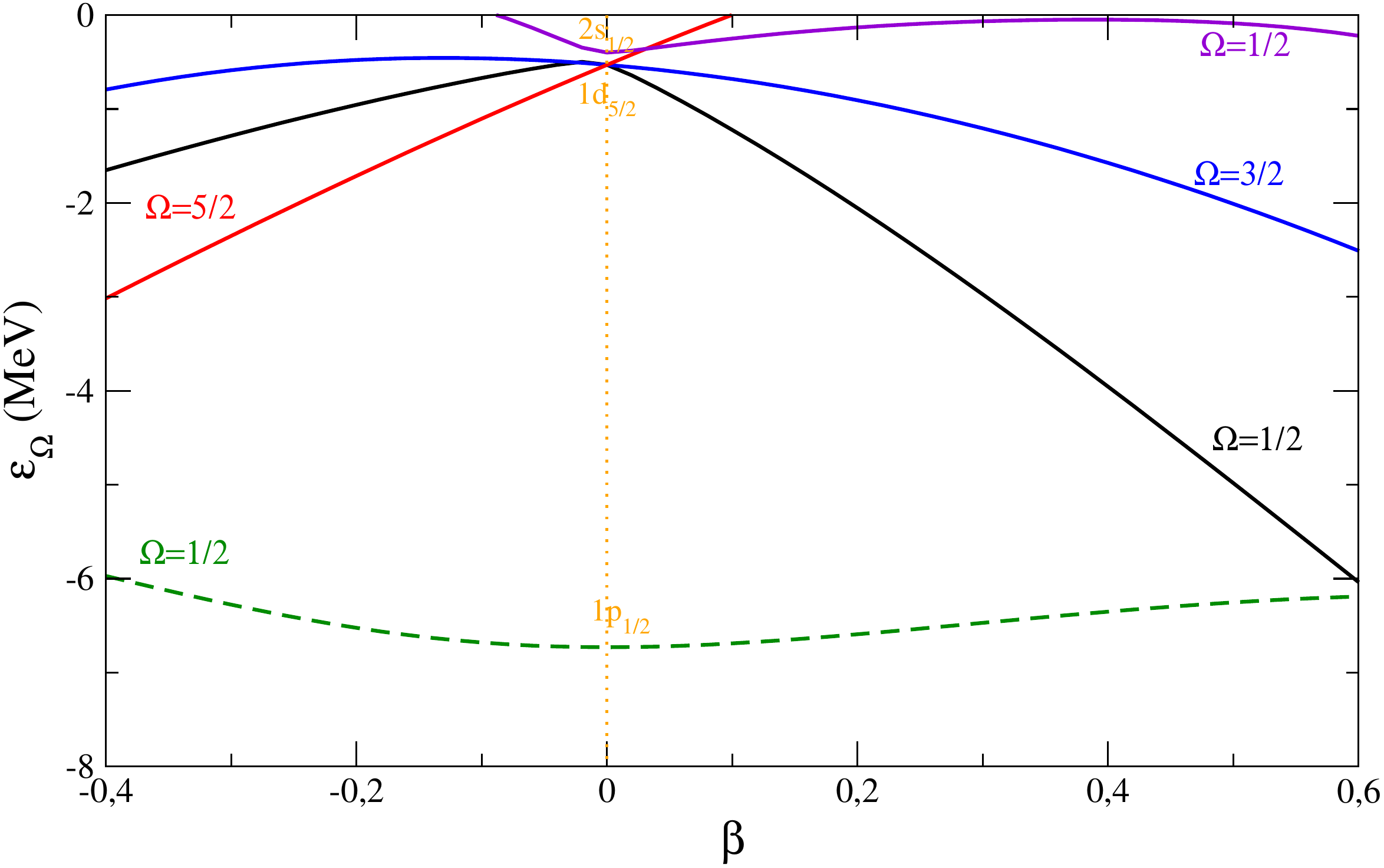}
\caption{\label{fig:NilssonDiagram} 
Nilsson Diagram obtained for $^{17}$C by diagonalization in the THO basis
using the parameters from \cite{Ham07}.
Solid lines represent positive parity levels, while
negative parity levels are represented by dashed lines.}
\end{figure}

Considering this potential, and the kinetic and  spin-orbit terms, 
with the parameters from \cite{Ham07} for $^{17}$C 
and different values of $\beta$,
the Nilsson diagram shown in Fig.~\ref{fig:NilssonDiagram} is obtained.
This diagram shows how the single-particle energy levels
of the valence nucleon change
as the deformation parameter varies.
Except for $\beta=0$, these levels do not have well-defined values of $\ell$ and $j$,
but they can be characterized by their parity $\pi$
and the projection $\Omega$ of $\vec j$ along the axial symmetry axis.
With the deformation, the spherical levels degenerated in $\Omega$ 
separate according to their $\Omega$ values 
and begin to mix with other levels with the same $\Omega$.
For $\beta>0$, the Nilsson levels originating from the same spherical level 
split in energy according to their $\Omega$ value, with higher $\Omega$ values lying at higher energies .
Due to the symmetry of the system,
the $\Omega$ and $-\Omega$ projection states are equivalent,
therefore each Nilsson level has twofold degeneracy. 

The complete Hamiltonian of the system also includes an intrinsic core term.
In this case, the core is approximated by a perfect rotor.
Therefore, $h_{core}$ depends on the the angular momentum of the core,
but the relation $\vec I=\vec J-\vec j$ can be used, resulting:
\begin{equation}
 h_{core}^{\text{Nilsson}}=
 \frac{\hbar^2}{2\mathcal J}\vec I^2=
 \frac{\hbar^2}{2\mathcal J}(\vec J-\vec j)^2,
\end{equation}
where $\mathcal J$ is the moment of inertia of the core.
This collective term mixes the different single-particle Nilsson levels, 
causing $\Omega$ to stop being a good quantum number
for the eigenstates of the full Hamiltonian.

In this model, the eigenstates are expressed as
\begin{equation}\label{eq:IntWF}
\Psi_{\varepsilon M}^{J^\pi}(\vec{r'},\omega)=
\sum_{\nu }R^{J^\pi}_{\varepsilon\nu }(r)\Phi_{\nu J}^M(\hat{r}',\omega),
\end{equation}
where now $\nu $ corresponds to the quantum numbers $\{\ell,s,j,\Omega\}$
with $\Omega>0$.
The functions $\Phi_{\nu J}^M(\hat{r}',\omega)$ are defined as
\begin{eqnarray}
\Phi_{\nu J}^M(\hat{r}',\omega)=
\frac{\sqrt{2J+1}}{4\pi}\left[
{\cal Y}_{\ell s}^{j\Omega}(\hat{r}')
{\cal D}_{M\Omega}^{J}(\omega)^*\right.+\nonumber\\
\left. (-1)^{J-j}
{\cal Y}_{\ell s}^{j-\Omega}(\hat{r}'){\cal D}_{M-\Omega}^{J}(\omega)^*\right].
\label{eq:IntAng}\end{eqnarray}
The definition of \cite{BS} is used for the
rotation matrices ${\cal D}_{M\Omega}^{J}(\omega)$
and the three Euler angles are denoted by $\omega$.
The functions $\Phi_{\nu J}^M(\hat{r}',\omega)$
are orthonormal and take into account the symmetry 
regarding the $\Omega$ and $-\Omega$ projections.

Assuming the core to be a rotor,
expressions (\ref{eq:LabWF}) and (\ref{eq:IntWF})
are equivalent and can be interchanged using the relation
\begin{equation}
R^{J^\pi}_{\varepsilon\alpha}(r)=
\sqrt{\frac{2I+1}{2J+1}}\sqrt{1+(-1)^I}
\sum_\Omega\langle j\Omega I 0|J\Omega\rangle
R^{J^\pi}_{\varepsilon\nu }(r),
\end{equation}
where $\ell$ and $j$ are the same for $\alpha$ and $\nu $
and $\langle j\Omega I 0|J\Omega\rangle$
is a Clebsch-Gordan coefficient.
This expression is obtained by transforming 
$\Phi_{\alpha J}^M(\vec{r},\xi)$ into 
$\Phi_{\nu J}^M(\hat{r}',\omega)$,
using the properties of the rotational matrices.

\subsection{\label{sec:level2}The THO Basis}
In this subsection, we briefly review the method followed in the present work to obtain the eigenvalues of the Hamiltonian and their associated wavefunctions.

The eigenstates of a two-body Hamiltonian,
like that of Eq.~(\ref{eq:H}),
follow the expression (\ref{eq:LabWF}) or (\ref{eq:IntWF}),
and the radial functions $R^{J^\pi}_{\varepsilon\alpha}(r)$
or $R^{J^\pi}_{\varepsilon\nu }(r)$
can be determined in several ways.
A common procedure is to insert the expansion
(\ref{eq:LabWF}) into the Schr\"odinger equation,
giving rise to a set of coupled differential equations
for the radial functions $R^{J^\pi}_{\varepsilon\alpha}(r)$
(see, e.g., \cite{BM}).

Alternatively, these functions can be obtained by
diagonalizing the Hamiltonian in a discrete basis. 
This basis is chosen in the form
$\psi_{n\tau JM}^{basis}=
R^{basis}_{n\ell}(r)\Phi_{\tau J}^{M}$,
where $\tau$ can be $\alpha=\{\ell,s,j,I\}$ or
$\nu =\{\ell,s,j,\Omega\}$ and the function $\Phi_{\tau J}^{M}$
is given by (\ref{eq:LabAng}) or (\ref{eq:IntAng}) respectively.
Thus, the eigenstates of the Hamiltonian can be expanded in the discrete basis 
as
\begin{equation}\label{eq:BasisEx}
\Psi_{iM}^{J^\pi}=
\sum_{n\tau}C_{n\tau}^{iJ^\pi}
\psi_{n\tau JM}^{basis}=
\sum_{n\tau}C_{n\tau}^{iJ^\pi}
R^{basis}_{n\ell}(r)\Phi_{\tau J}^{M}.
\end{equation}

There are many possible choices for
the basis functions $R^{basis}_{n\ell}(r)$ (Gaussian, harmonic oscillator,
Laguerre, etc.). In this work we use the transformed
harmonic oscillator (THO) basis, obtained from the harmonic oscillator
basis with an appropriate local scale transformation (LST) \cite{SP88,PS91}.

If the LST function is  
denoted by $s(r)$, the THO states are obtained as: 
\begin{equation}
\label{eq:tho}
R ^{THO}_{n\ell}(r)=\frac sr\sqrt{\frac{ds}{dr}} R^{HO} _{n\ell}[s(r)],
\end{equation}
where $R ^{HO}_{n\ell}(s)$ (with $n = 1,2,...$) is the radial part of the usual HO functions. 
According to the definition given above, the LST is indeed not unique.
Here, we adopted a parametric form for the LST from Karataglidis~\textit{et al.}~\cite{Amos} 
\begin{equation}
\label{lstamos}
s(r)  =
 \left[  \frac{1}{   \left(  \frac{1}{r}
    \right)^m  +  \left( 
\frac{1}{\gamma\sqrt{r}} \right)^m } \right]^{\frac{1}{m}}\ ,
\end{equation}
that depends on the parameters $m$ and $\gamma$. The extension of $R ^{HO}_{n\ell}(s)$ will also depend on the oscillator length
$b$. Note that,  
asymptotically, the function  $s(r)$ behaves as 
$s(r)\sim \gamma \sqrt{r}$
and hence the functions obtained by applying this LST to the HO basis
behave at  
large distances as $\exp(-\gamma^2 r / 2 b^2)$.
Therefore, the ratio
$\gamma/b$ can be related to an effective linear momentum, 
$k_\mathrm{eff}=\gamma^2 /2b^2$, which  
governs the asymptotic behavior of the THO functions. As the ratio
$\gamma/b$ increases, the radial extension of the basis decreases and,  
consequently, the eigenvalues obtained upon diagonalization of the
Hamiltonian in the THO basis tend to spread at higher excitation
energies. Therefore, $\gamma/b$ determines the density of eigenstates
as a function of the excitation energy. 
In all calculations
presented in this work, the power $m$ has been taken as $m=4$. This
choice is discussed in  Ref.~\cite{Amos} where the authors found that
the results are weakly dependent on $m$. 

Note that, by construction, the family of functions  
\( R ^{THO}_{n\ell}(r) \) constitute a complete orthonormal set.
Moreover, they decay exponentially
at large distances, thus ensuring the correct asymptotic behaviour
for the bound wave functions. 
In practical calculations, a finite set of $\tau$ channels
and wavefunctions as in Eq. (\ref{eq:tho})
are retained, and the Hamiltonian is
diagonalized in this truncated basis,  
giving rise to a set of eigenvalues $\left\{\varepsilon_i^{J^\pi} \right\}$
and their associated
eigenfunctions, $\left\{\Psi_{iM}^{J^\pi}\right\}$. As the basis size
is increased,  the eigenstates with negative energy
will tend to the exact bound states of the system, while those
with positive eigenvalues can be regarded as a finite representation
of the unbound states. 

This analytical THO basis has been successfully applied to the structure and reactions of two-body systems in \cite{Mor09} and generalized to the case in which core excitations are included \cite{Lay12}.

\section{\label{sec:level1}One Neutron Transfer Reactions}

As in the case of stable nuclei, a significant source of information of halo nuclei stems from the analysis of transfer reactions involving these nuclei.
We focus on $C(d,p)A$ and $A(p,d)C$ reactions,
where $A$ corresponds to one of the nuclei studied with our models and $C$ to its respective core.
In this case, they are studied using the
adiabatic distorted wave approximation (ADWA) \cite{ADWA}.
The formalism for this approximation is identical to that of
the distorted wave Born approximation (DWBA),
with the difference that adiabatic potentials are calculated
between deuteron and the other nucleus.
The reason for using ADWA instead of DWBA is to take into account, 
approximately, the effect of deuteron break-up on the calculation.

Considering for defiteness the $(d,p)$ case, the transition amplitudes are calculated in \textit{post} form
\begin{equation}\label{Tpost}
\mathcal{T}_{if}^{post}=
\langle\chi_{\vec k_{pA}}^{(-)}\psi_{CA}
|V_{pn}+U_{pC}-U_{pA}|
\chi_{\vec k_{dC}}^{(+)}\psi_{d}\rangle,
\end{equation}
where $\chi_{\vec k_{dC}}$ and $\chi_{\vec k_{pA}}$ are distorted waves for the entrance and exit channels, respectively, depending on the corresponding deuteron and proton momenta. The function $\psi_{d}$ stands for deuteron ground state wavefunction, generated with the potential $V_{pn}$. The operators $U_{pC}$ and $U_{pA}$ are optical potentials for the $p+C$ and $p+A$ systems.
Our structure model is implemented in the overlap $\psi_{CA}\equiv\langle C|A\rangle$.
Starting from expression~(\ref{eq:LabWF}), it can be shown
\begin{equation}\label{overlap}
\psi_{CA}(\vec r)=\sum_j\langle JM|jm_jIm_I\rangle
R_{\varepsilon\alpha}^{J^\pi}(r)
\mathcal{Y}_{\ell s}^j(\hat r).
\end{equation}
Therefore, only the $R_{\varepsilon\alpha}^{J^\pi}(r)$
functions resulting from our models are needed.
The transition amplitudes are calculated for given states of nuclei $A$ and $C$,
in our models, which implies well-defined $\{J^\pi,\varepsilon,I\}$ values with their compatible $j$ values.

Similarly, in the $(p,d)$ case, the transition amplitudes are calculated in \textit{prior} form.
The amplitudes and their corresponding cross sections are calculated using FRESCO code \cite{fresco}.

\section{\label{sec:level1}Application to  $^{\textbf{17}}$C
and $^{\textbf{11}}$Be}
The Nilsson Hamiltonian has been built
and diagonalized in the THO basis
for the $^{17}$C and $^{11}$Be systems.
To ensure convergence,
the values $0\leq\ell\leq6$ and $1\leq n\leq30$ have been considered.
The results of this model have been compared with
those obtained with the PAMD model from \cite{Lay12},
also diagonalized in the THO basis.
In this case, $0\leq\ell\leq3$ and $1\leq n\leq30$ are used.
This model only considers the ground state $0^+$ and
first excited state $2^+$ for the core, therefore
it would not be consistent to consider higher values of $\ell$.

\subsection{\label{sec:level2} Structure of $^{\textbf{17}}$C}
In the Nilsson calculations presented in this work,
the geometry of the central Woods-Saxon potential used in \cite{Ham07} for $^{17}\text{C}$ 
is adopted ($R=3.266 \text{ fm}$, $a=0.67\text{ fm}$).
The strength of this central potential $V_c(r)$ is fixed to $44.27\text{ MeV}$.
Consequently, following the relation from~\cite{Hamamoto05},
a $8.825\text{ MeV}$ strength is obtained
for the spin-orbit part $V_{\ell s}(r)$ keeping the same geometry.
The deformation parameter $\beta$ takes the value 0.34,
similar to the value 0.33 used by Amos \textit{et al.} \cite{AMOS12},
and for the core Hamiltonian $\hbar/2\mathcal{J}=0.3\text{ MeV}$ is used,
compatible with the excitation energy of the first excited state
$2^+$ of $^{16}\text{C}$ ($1.766\text{ MeV}$ \cite{TILLEY93}).

\begin{figure}
\includegraphics[width=0.9\linewidth]{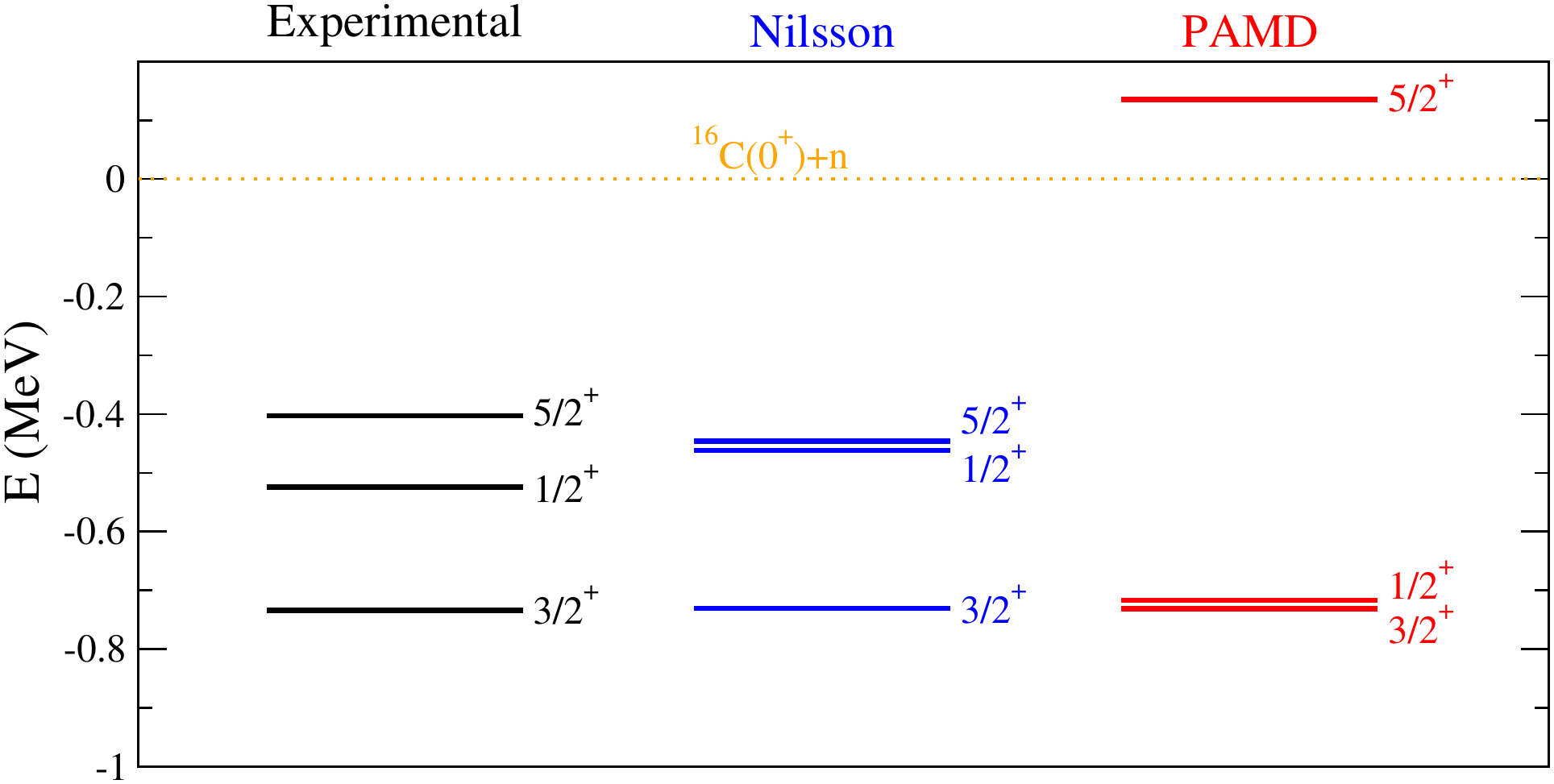}
\caption{\label{fig:levels_c17} Experimental and calculated energy levels of $^{17}$C.
Starting from the left, the second column is the Nilsson model and the third the PAMD.
Experimental values are from \cite{Wang_2017,Ele05}.}
\end{figure}

With these parameters, the Hamiltonian is fully defined and diagonalized on the THO basis
using $b=2.4\text{ fm}$ and $\gamma=2.7\text{ fm}^{1/2}$.
Thus, the energies of the bound states of \textsuperscript{17}C,
shown in the central spectrum of Fig.~\ref{fig:levels_c17},
are negative eigenvalues of this Hamiltonian.
In the same figure, on the left,
the experimental data \cite{Wang_2017,Ele05} for these levels are shown and,
 on the right, the results of the PAMD model.
In all cases we have a $3/2^+$ ground state,
a $1/2^+$ first excited  and a  $5/2^+$ second excited state.
It should be noted that while the PAMD model predicts
the second excited state as a near-threshold resonance,
in the new Nilsson model it appears as a bound state
whose energy is closer to the experimental one.
Also, the first excited state is closer in energy
to the experimental value in the Nilsson model than in the PAMD.

\begin{figure}
\includegraphics[width=0.9\linewidth]{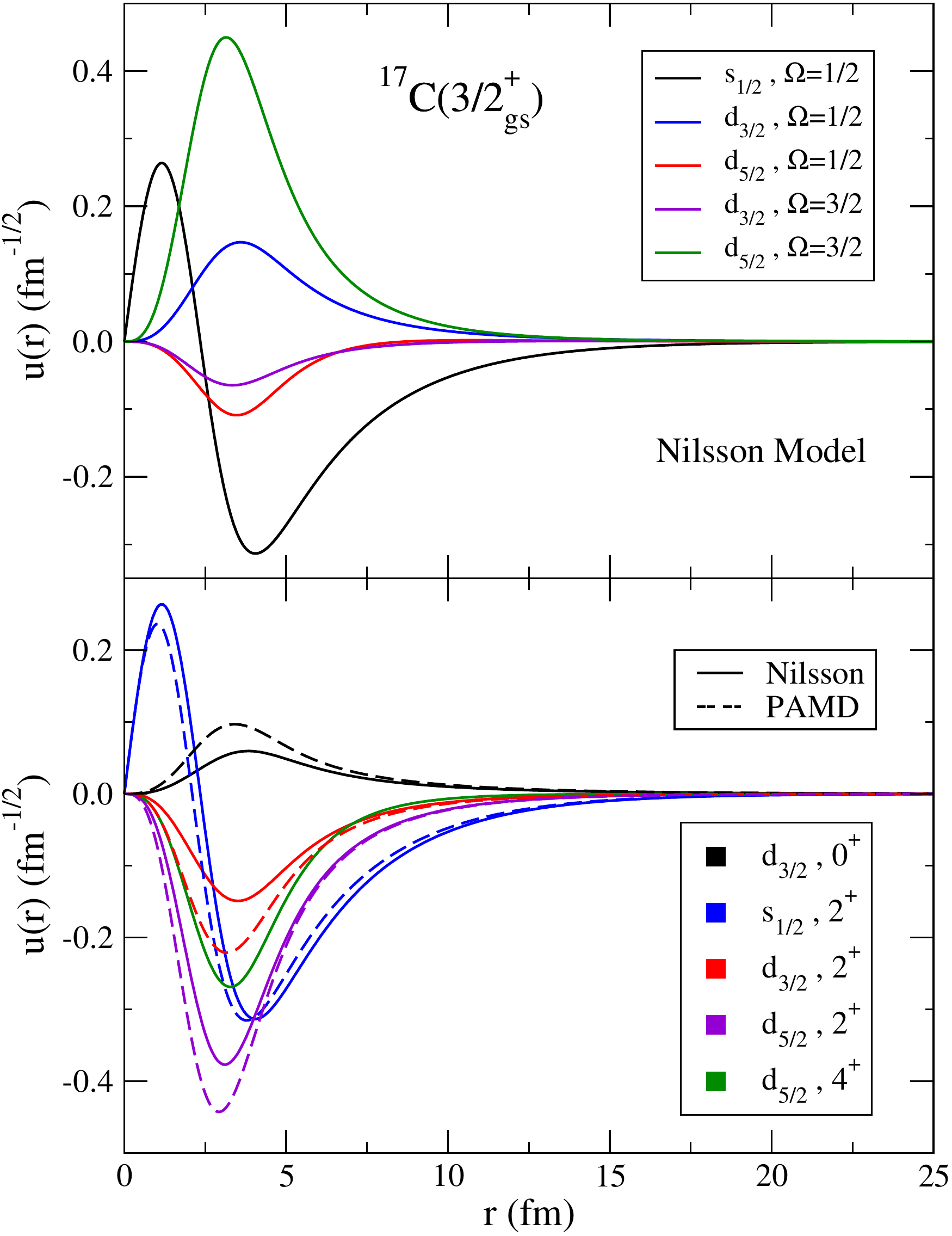}
\caption{\label{fig:wf_c17_gs} Radial part of the wave function
obtained for the ground state of $^{17}$C.
The upper panel shows the most relevant components in the Nilsson Model 
according to their quantum numbers $\ell$, $j$ and $\Omega$.
The lower panel shows the components according to $\ell$, $j$ and $I$.}
\end{figure}

The radial parts of the ground state  wave function are shown in Fig.~\ref{fig:wf_c17_gs}.
The top panel shows the functions 
$u_{\nu }(r)=rR^{J^\pi}_{\varepsilon\nu }(r)$
for the Nilsson model,
whereas on the lower panel the functions $u_{\alpha}(r)=rR^{J^\pi}_{\varepsilon\alpha}(r)$ are compared for both models.
The resulting functions differ mainly in the norm of the components,
which are the weights of Table \ref{tab:c17_weights}.
In our assumed simplified two-body model,
in which antisymmetrization between the valence neutron and the core is neglected,
spectroscopy factors cannot be strictly obtained.
However, as long as these antisymmetrization effects are not large,
these weights can be approximately regarded as SF.
It should be noted that the Nilsson model predicts a
non-negligible weight for the component with a $4^+$ core state,
while the PAMD model does not consider this core state.
This means that, in general, 
the rest of the weights are lower in the Nilsson model.

\begin{table}
\caption{\label{tab:c17_weights}%
Weight of the main components for the wave function of 
the ground state and first excited of \textsuperscript{17}C.
Components with weights less than 0.005 are not included.}
\begin{ruledtabular}
\begin{tabular}{ccccc}
State &\multicolumn{2}{c}{$3/2^+_{\text{gs}}$}&
\multicolumn{2}{c}{$1/2^+_1$}\\
Model&Nilsson&PAMD&Nilsson&PAMD\\
\hline
$|(\ell s)j\otimes0^+\rangle$& 0.012 & 0.028 & 0.668 & 0.512\\
$|s_{1/2}\otimes2^+\rangle$& 0.375 &  0.349 & - & -\\
$|d_{3/2}\otimes2^+\rangle$& 0.064 & 0.131 & 0.028 & 0.040\\
$|d_{5/2}\otimes2^+\rangle$& 0.366 & 0.492 & 0.299 & 0.448\\
$|d_{5/2}\otimes4^+\rangle$& 0.179 & - & -& -\\
\end{tabular}
\end{ruledtabular}
\end{table}

\begin{figure}
\includegraphics[width=0.9\linewidth]{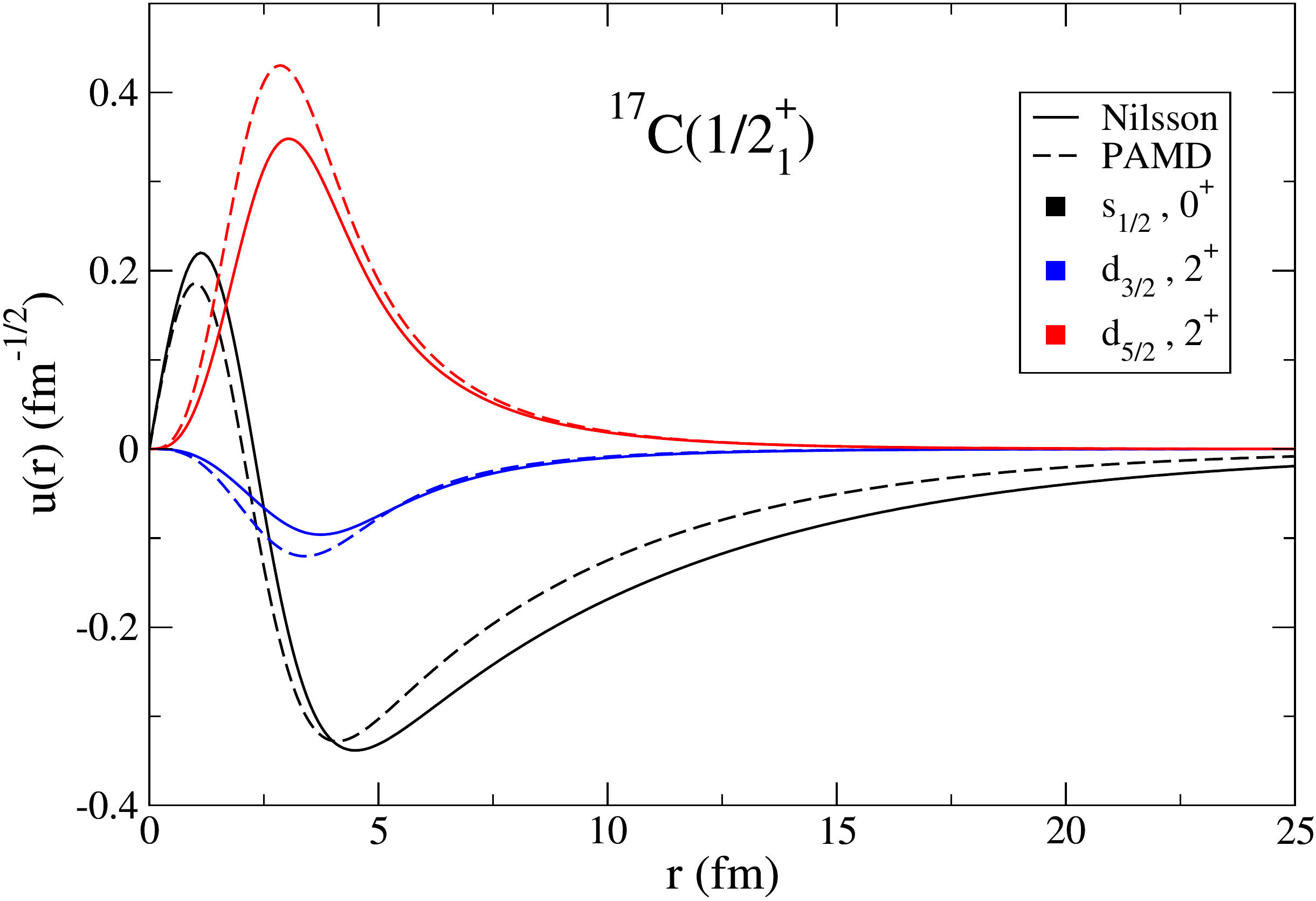}
\caption{\label{fig:wf_c17_1ex} Radial part of the wave function
obtained for the first excited state of $^{17}$C.}
\end{figure}

Figure \ref{fig:wf_c17_1ex} shows the radial parts of the first excited state wave function
of $^{17}\text{C}$.
It can be seen that the spatial extension is much larger in this case.
The mean square radius for this state is 6.55~fm
in the Nilsson model and 5.24~fm in the PAMD,
while for the ground state the value is around 4~fm in both models.
Therefore, the results of both models corroborate the halo nature of this state. 
Table \ref{tab:c17_weights} also compares the weights of the components for this state.

\subsection{\label{sec:level2}Structure of $^{\textbf{11}}$Be}
Our Nilsson Hamiltonian for \textsuperscript{11}Be consists of a central Woods-Saxon potential ($R=2.483 \text{ fm}$ $a=0.65 \text{ fm}$), a spin-orbit potential with strength $8.5\text{ MeV}$
and the deformation parameter ($\beta=0.67$) from \cite{Lay12}.
We use $\hbar/2\mathcal{J}=0.56\text{ MeV}$ for the $^{10}\text{Be}$ core,
which is compatible with the excitation energy of
its first $2^+$ state ($3.368\text{ MeV}$ \cite{TILLEY04}).
For the central potential, a similar parity-dependent strength is used 
($52.43\text{ MeV}$ for positive-parity states and $49.62\text{ MeV}$ for negative ones).
In the same way, the PAMD model allows for a parity-dependent renormalization factor.
The purpose of this parity dependence is to reproduce the inversion
of the $1/2^+$ and $1/2^-$ bound state levels of  \textsuperscript{11}Be.
As explained in \cite{Lay14},
this inversion is partly ascribed to core deformation,
but also to other effects not included in our treatment.
In any case, the difference between strengths according
to parity does not exceed 6\% for both models.

\begin{figure}
\includegraphics[width=0.9\linewidth]{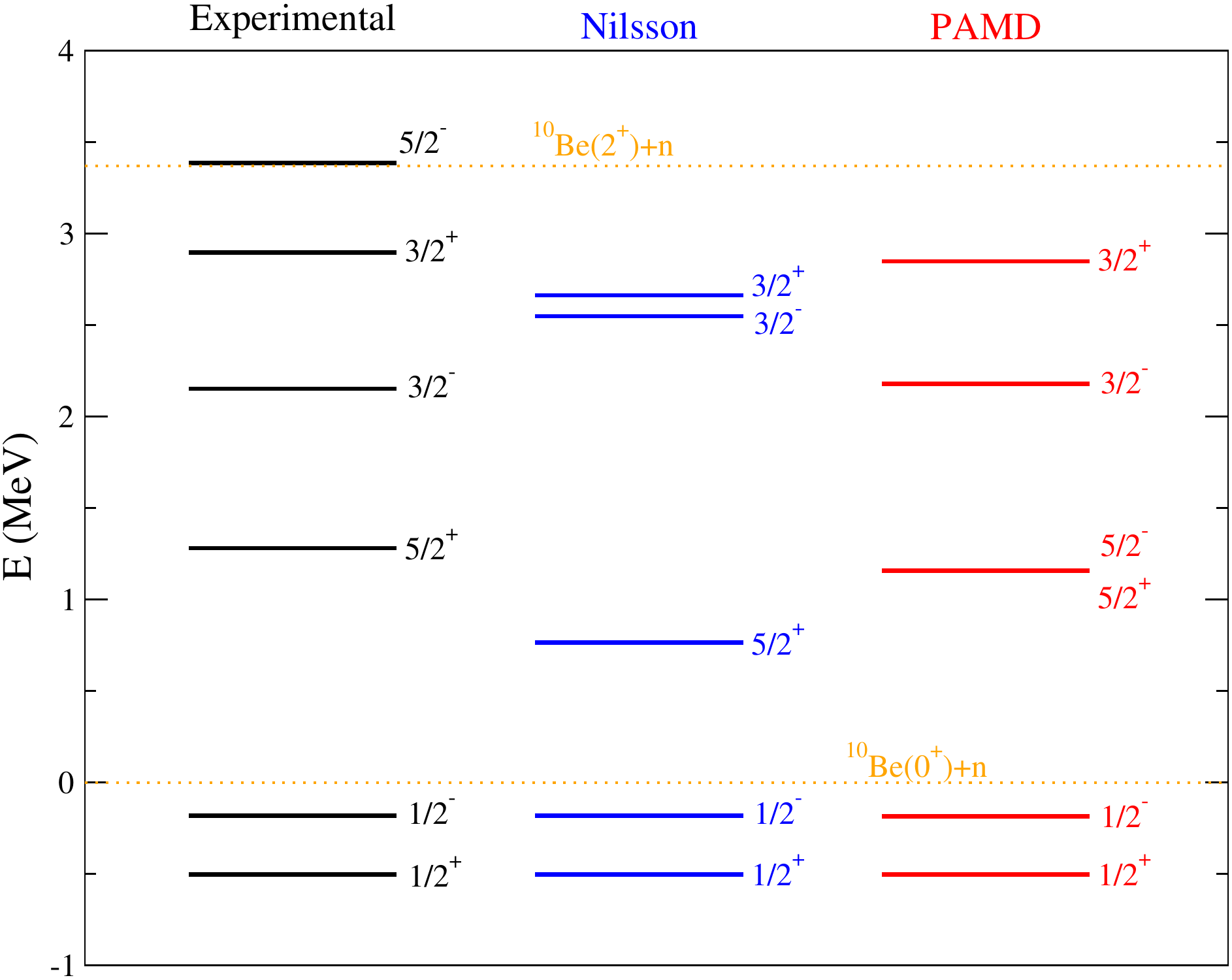}
\caption{\label{fig:levels_be11} 
Spectrum obtained for $^{11}$Be with the Nilsson and the PAMD models 
calculations compared with the experimental one \cite{KELLEY2012,Fuk04}.}
\end{figure}

The Nilsson Hamiltonian is diagonalized in a THO basis
with $b=2.0$ and $\gamma=2.5\text{ fm}^{1/2}$.
Figure \ref{fig:levels_be11} compares the \textsuperscript{11}Be states
obtained with this model up to 4 MeV, those obtained with PAMD
and the experimental levels \cite{KELLEY2012,Fuk04}.
In this figure, in addition to the $^{10}\text{Be}(0^+)+n$ threshold,
the $^{10}\text{Be}(2^+)+n$ threshold is also indicated.
Both models reproduce the experimental energies of the ground state $1/2^+$ and
the first excited state $1/2^-$, while the energies of the resonances $5/2^+$, $3/2^-$
and $3/2^+$ are better reproduced in the PAMD model.
In case of the $5/2^-$ resonance, the Nilsson model predicts it above 4 MeV
and, for the PAMD model, it is well below.
The reason for this difference is the effect of
including the $4^+$ core state in the calculation.
In the PAMD model, the state $5/2^-$ corresponds mostly to a
$|\text{p}_{1/2}\otimes2^+\rangle$ configuration.
However, in the Nilsson model, there is a mixture of configurations
$|\text{p}_{3/2}\otimes4^+\rangle$ and $|\text{p}_{1/2}\otimes2^+\rangle$
resulting in two possible  $5/2^-$ resonances.
One of them, for which the component with $\text{p}_{3/2}$ dominates,
is close to the threshold $^{10}\text{Be}(0^+)+n$,
but it is a forbidden state due to the Pauli exclusion principle.
The other is allowed, but it is above 4 MeV.

\begin{figure}
\includegraphics[width=0.9\linewidth]{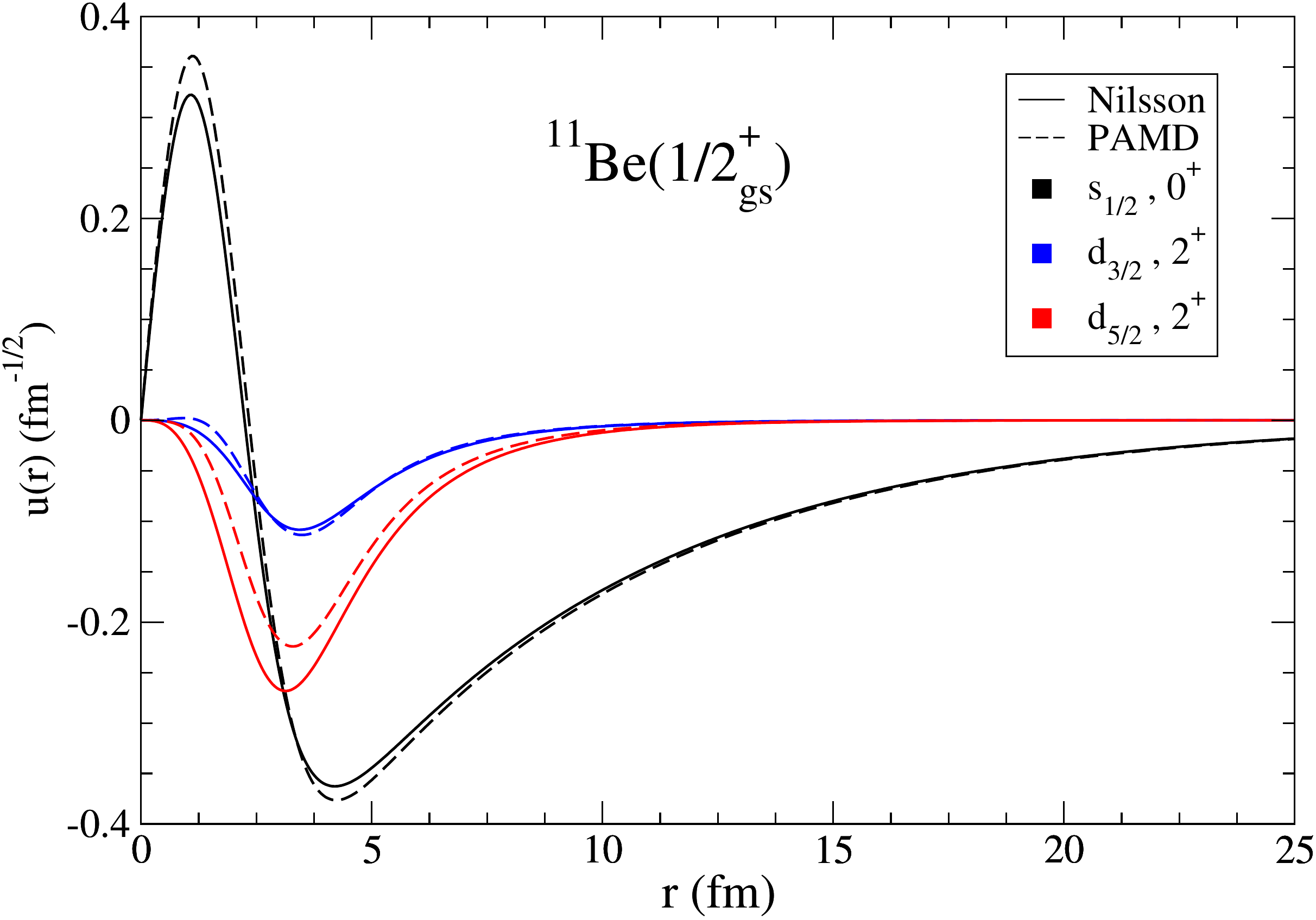}
\caption{\label{fig:wf_be11_gs} Radial parts of the ground state wave function
of $^{11}$Be.}
\end{figure}

\begin{figure}
\includegraphics[width=0.9\linewidth]{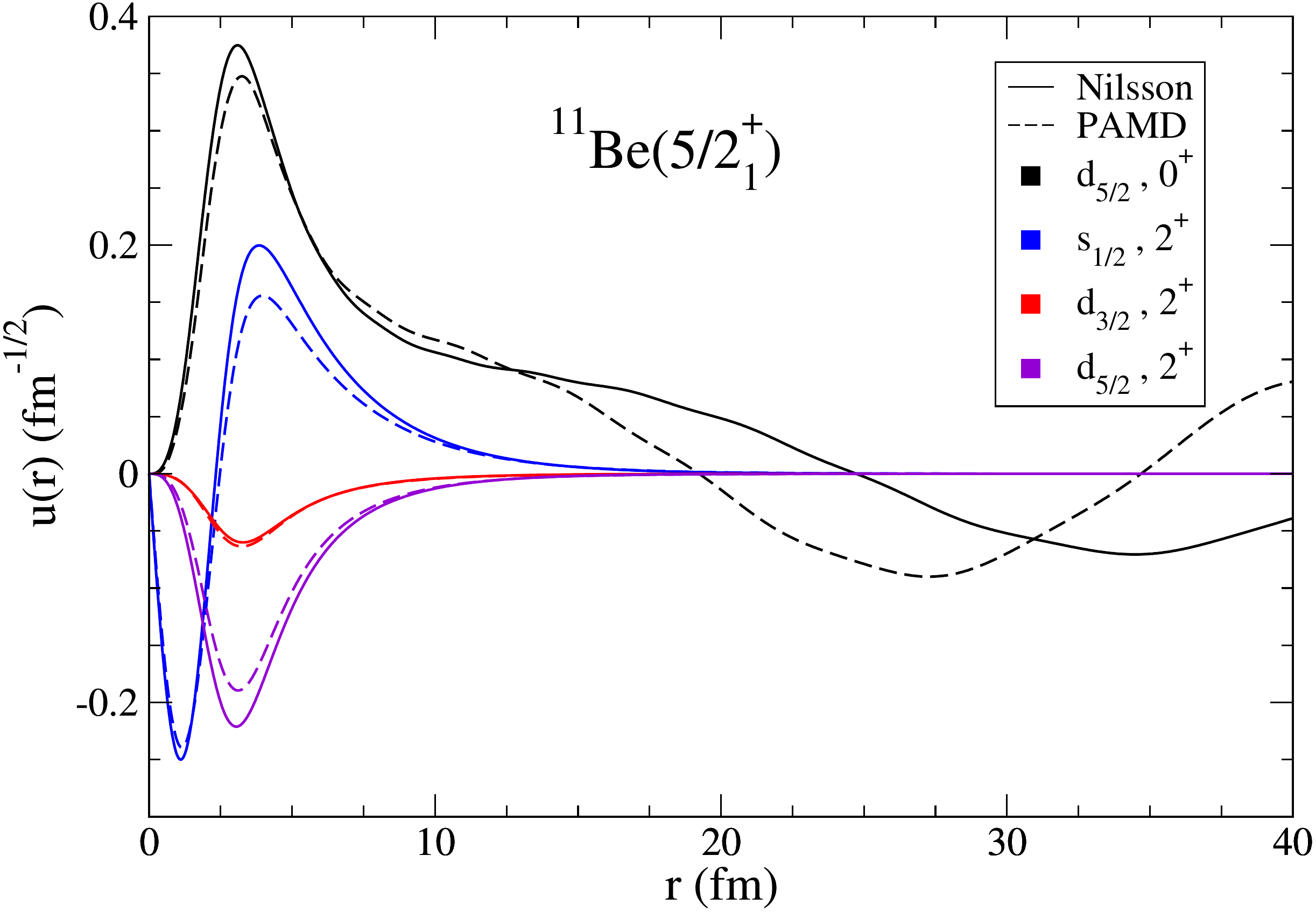}
\caption{\label{fig:wf_be11_res} Radial part of the
continuum wave function for the first $^{11}$Be resonance.
The Nilsson wave function corresponds to an energy of 0.77 MeV
above  the $^{10}\text{Be}(0^+)+n$ threshold,
while the result of the PAMD model is for 1.14 MeV.}
\end{figure}

Figure \ref{fig:wf_be11_gs} compares the wave function
of the \textsuperscript{11}Be ground state obtained 
for the Nilsson (the most relevant components) and PAMD models.
The results are quite similar, 
with a clear dominance of the $|s_{1/2}\otimes0^+\rangle$ component.
It can be seen that this component has a greater weight in the PAMD model,
while the opposite occurs with the $|d_{5/2}\otimes2^+\rangle$ component.
Clearly, the $|d_{3/2}\otimes2^+\rangle$ component is not very relevant for this state.
Likewise, Fig.~\ref{fig:wf_be11_res} shows the radial part
of the wave function for the first $5/2^+$ resonance  obtained in both models.
Since the resonances occur at different energies depending on the model,
the asymptotic part is clearly different.
However, the behavior is very similar for $r<5\text{ fm}$.

\section{\label{sec:level1} Application to reactions}

The differential cross sections for the transfer reactions
$^{16}\text{C}(d,p)^{17}\text{C}$ and $^{11}\text{Be}(p,d)^{10}\text{Be}$
have been calculated using the adiabatic distorted wave approximation (ADWA) \cite{ADWA}.
The \textit{post} and the \textit{prior} form are used respectively,
which require the overlap functions $\langle ^{17}\text{C}|^{16}\text{C}\rangle$
and $\langle ^{10}\text{Be}|^{11}\text{Be}\rangle$.
These functions are taken from the results of our structure models.

\subsection{\label{sec:level2}
$^{\textbf{16}}\textbf{C}\bm{(d,p)}^{\textbf{17}}\textbf{C}$}

For this reaction with carbon, 
the cross section for the transfer to bound states of $^{17}$C has been calculated.
The results are compared with the recent experimental data from GANIL \cite{Pereira}.
These data were obtained in inverse kinematics with a $^{16}$C beam at $17.2$~MeV/nucleon.
Regarding the potentials used in the reaction calculation, the Chapel-Hill (CH89) parameterization \cite{Var91} was employed for the 
p+\textsuperscript{17}C optical potential,
whereas the Reid soft-core potential \cite{RSC} was used for the n+p interaction.
The d+\textsuperscript{16}C adiabatic potential was built
within the Johnson-Tandy finite-range prescription \cite{JT},
assuming the CH89 parameterization for the nucleon+\textsuperscript{16}C potentials.

\begin{figure}
\includegraphics[width=0.9\linewidth]{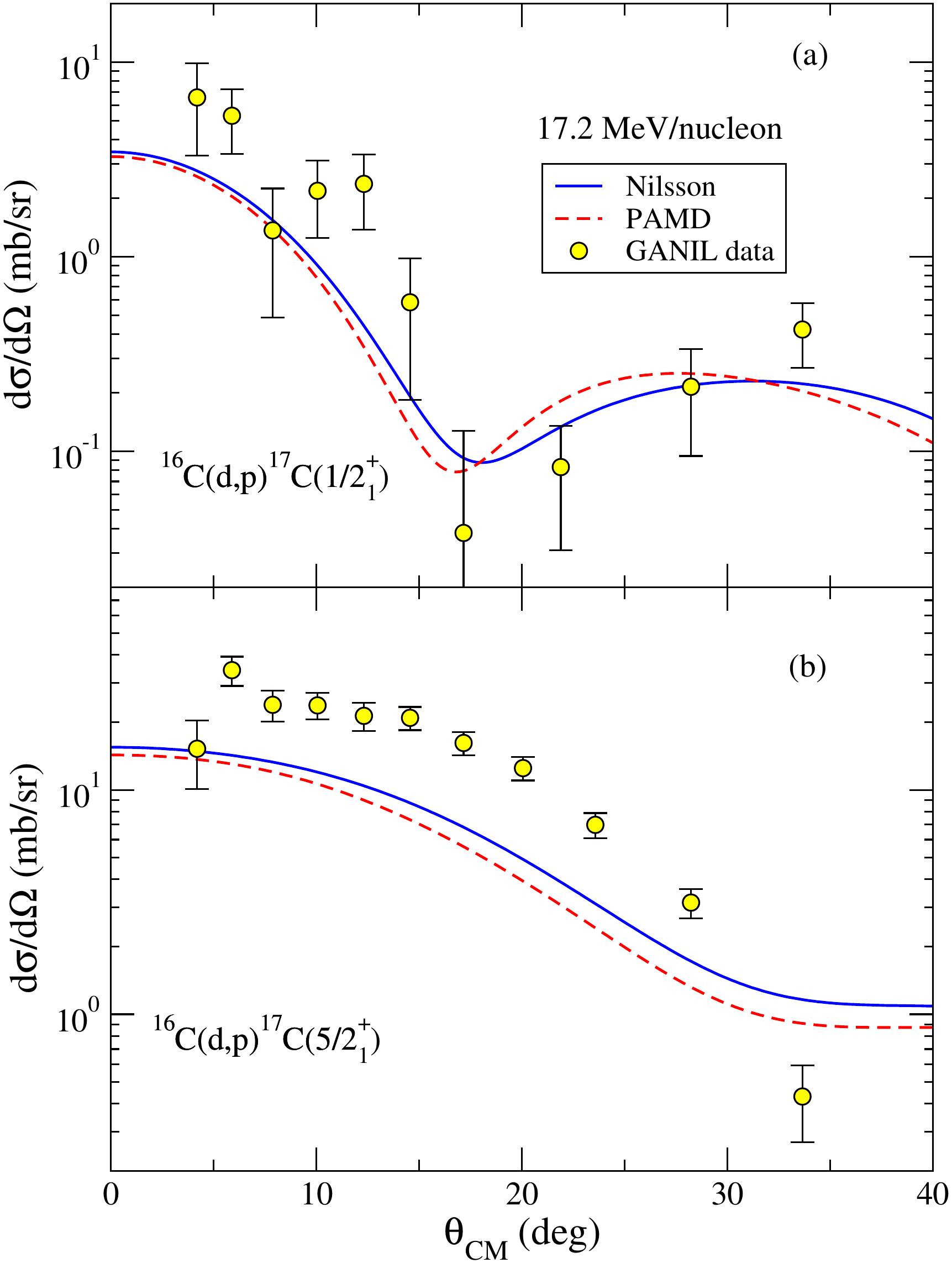}
\caption{\label{fig:c16dpc17_ex}  
Angular distribution of the $^{16}\text{C}(d,p)^{17}\text{C}$
reaction at 17.2 MeV/nucleon
when the $^{17}\text{C}$ first excited state $1/2^+_1$ (upper pannel),
and second excited state $5/2^+_1$ (lower pannel) are populated.
The results using the Nilsson and PAMD models are compared
with the experimental data \cite{Pereira}.}
\end{figure}

\begin{figure}
\includegraphics[width=0.9\linewidth]{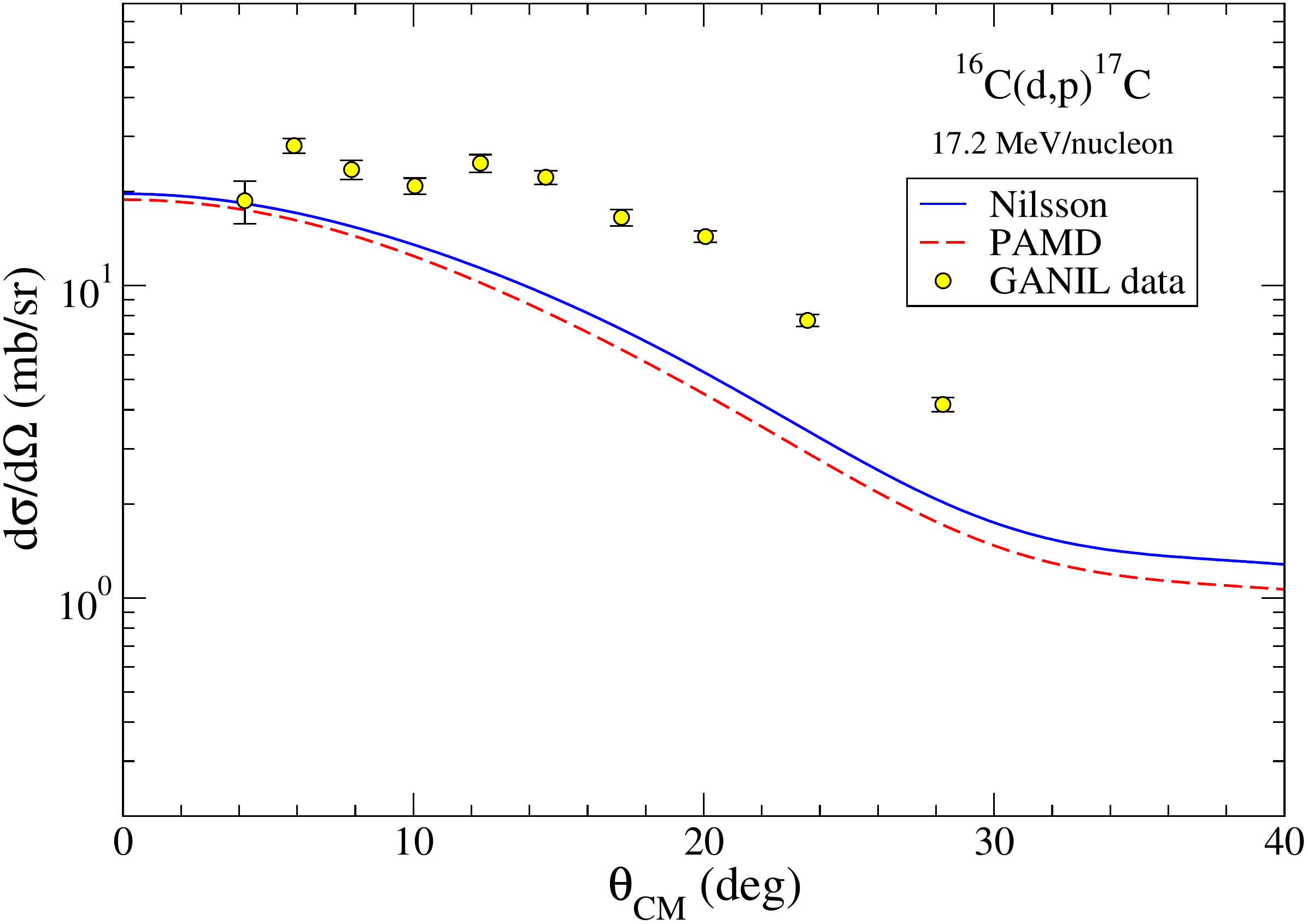}
\caption{\label{fig:c16dpc17_bound} 
Differential cross section of $^{16}\text{C}(d,p)^{17}\text{C}$
for transfer to bound states at 17.2 MeV/nucleon. For both models,
the sum of the results for each bound state are shown
and they are compared with the experimental data \cite{Pereira}.}
\end{figure}

Figure \ref{fig:c16dpc17_ex} shows the comparison
of the results of the two models with the experimental data
when the first and second excited states of \textsuperscript{17}C are populated.
For the case of the first excited state, we find good agreement between the results of our models and the data
 (Fig.~\ref{fig:c16dpc17_ex}(a)).
From the comparison of the experimental data 
with finite-rage ADWA calculations using the CH89 parameterization,
spectroscopy factors (SF) are obtained in Ref.~\cite{Pereira}.
A value of $0.80\pm0.22$ is shown for the configuration 
$|s_{1/2}\otimes0^+\rangle$ of the \textsuperscript{17}C $1/2^+_1$ state.
This SF is compatible with the 0.67 obtained with the Nilsson model,
but not so much with the 0.51 of the PAMD (see table~\ref{tab:c17_weights}).
For the second excited state (Fig.~\ref{fig:c16dpc17_ex}.b),
both models obtain a reasonable result for the angular distribution,
although they underestimate the experimental cross section.
From these data, a $0.62\pm0.13$ spectroscopy factor is obtained
for the configuration $|d_{5/2}\otimes0^+\rangle$,
while the Nilsson and PAMD models obtain only 0.33 and 0.32 respectively.
To distinguish these two states experimentally 
it has been necessary to measure $\gamma$-rays in coincidence, 
therefore the angular distribution for the ground state is not presented. 
However, Fig.~\ref{fig:c16dpc17_bound} shows the angular distribution
when any of the bound states of \textsuperscript{17}C are populated,
that is, the sum of contributions from each of those states.
Furthermore, we can say that both the experimental data and our two models agree 
that the spectroscopic factor in this case
($|d_{3/2}\otimes0^+\rangle$) is well below 0.1.
Considering the results, transfer to second excited state 5/2$^+$
is by far the greatest contribution for this reaction.

\subsection{\label{sec:level2}
$^{\textbf{11}}\textbf{Be}\bm{(p,d)}^{\textbf{10}}\textbf{Be}$}

The differential cross section for this transfer reaction has been calculated 
when the ground state 0$^+$ and first excited state 2$^+$ of \textsuperscript{10}Be are populated.
Calculations have been performed for two different incident energies and
the results are compared with the experimental data corresponding to those energies.
For both energies, the calculations have been performed using the Johnson-Tandy prescription
for the d+\textsuperscript{10}Be potential \cite{JT} and the Reid Soft-Core n+p interaction \cite{RSC}.
Experimentally, in both cases the reaction has been studied in inverse kinematics using a $^{11}$Be beam.

\begin{figure}
\includegraphics[width=0.9\linewidth]{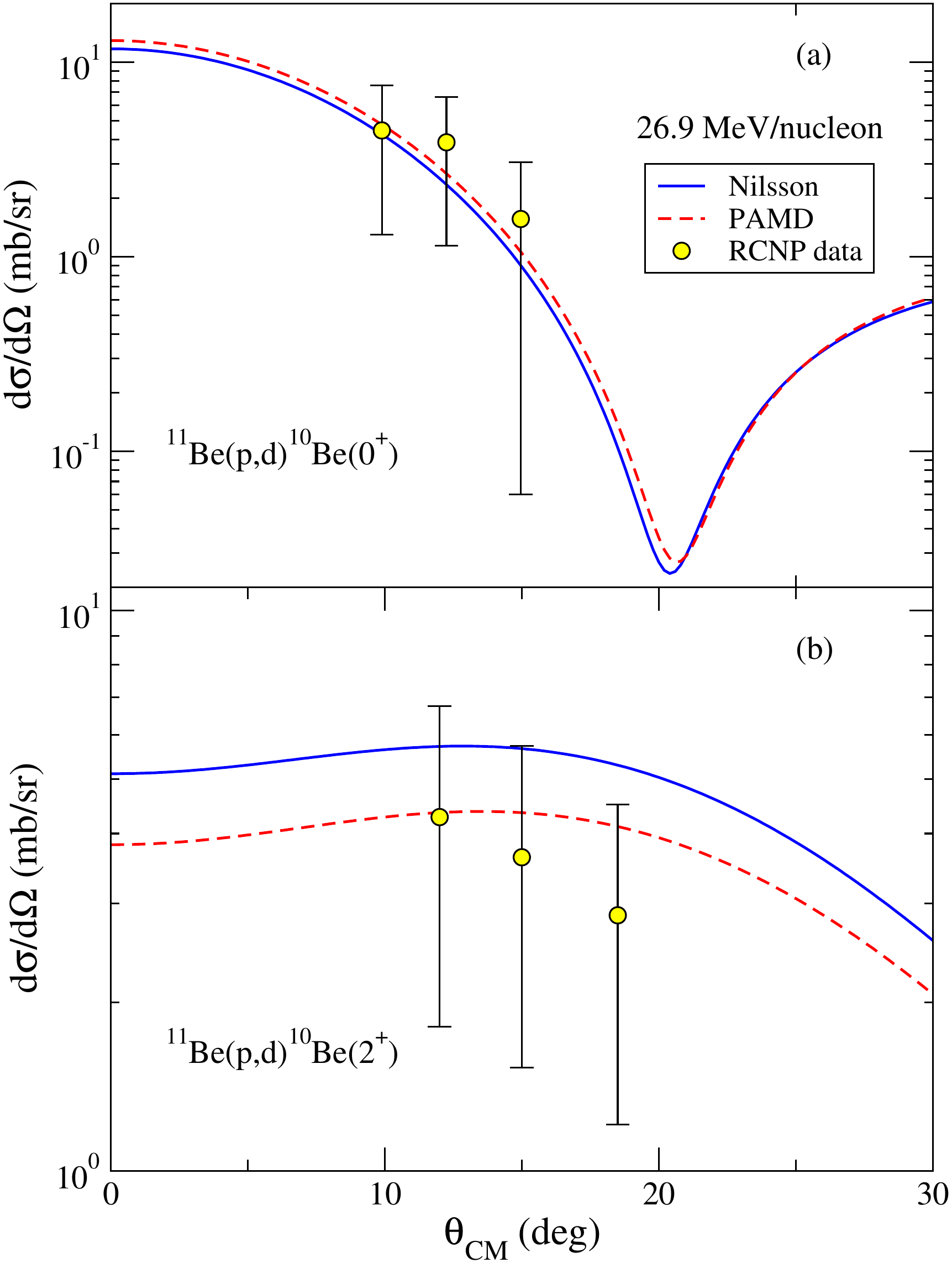}
\caption{\label{fig:be11pdbe10_1}  
Angular distribution of the $^{11}\text{Be}(p,d)^{10}\text{Be}$
cross section at 26.9 MeV/nucleon
when $^{10}\text{Be}$ ground state $0^+$ (upper pannel),
and first excited $2^+$ (lower pannel) are populated.
The solid lines represent the results applying the different models
and they are compared with the experimental data \cite{Jiang_2018}.
}
\end{figure}

First, the comparison of our calculations with the data 
from the Research Center for Nuclear Physics (RCNP)
for a $^{11}$Be beam 26.9 MeV/nucleon \cite{Jiang_2018}
is shown in Fig.~\ref{fig:be11pdbe10_1}.
For this calculation, a renormalized CH89 parameterization has been used
for the p+\textsuperscript{11}Be optical potential~\cite{CH89Chen}.
This potential applies two normalization factors
to the central part of the CH89 parameterization
to improve the agreement with the experimental data 
of the elastic scattering at 26.9 MeV/nucleon:
0.78 for the real part and 1.02 for the imaginary part.
CH89 parameterization was also employed
for the construction of the d+\textsuperscript{10}Be adiabatic potential.

Taking into account the error bars in Fig. \ref{fig:be11pdbe10_1},
the results of both models are compatible with the experimental data.
Panel (a) corresponds to the transfer to the ground state 0$^+$ of \textsuperscript{10}Be,
where the results of both models are extremely similar.
However, panel (b) shows that when the 
first excited state 2$^+$ is populated,
the PAMD model gives a smaller cross section,
getting closer to the data.

\begin{figure}
\includegraphics[width=0.9\linewidth]{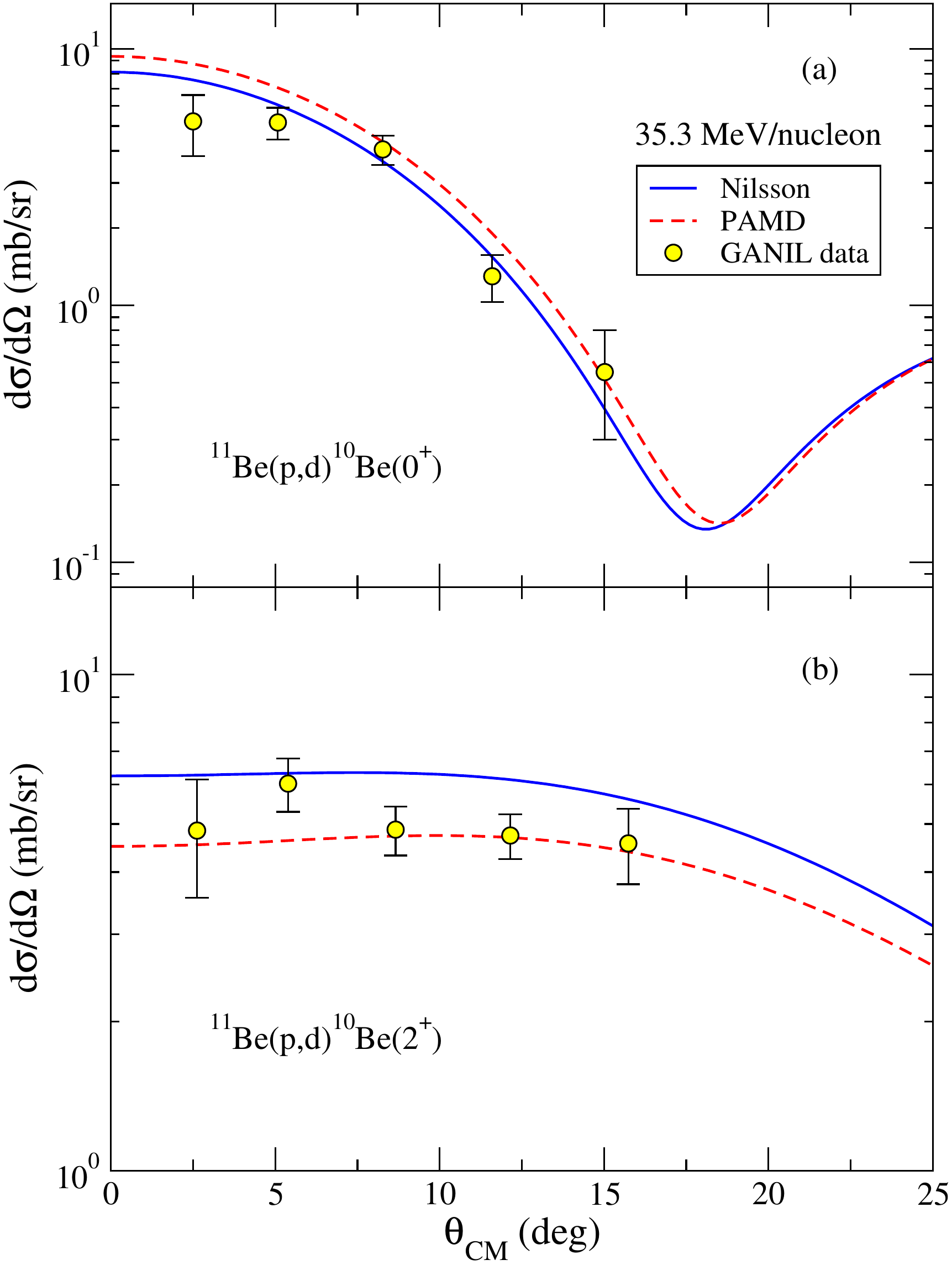}
\caption{\label{fig:be11pdbe10_2} Differential cross section for
the $^{11}\text{Be}(p,d)^{10}\text{Be}$ transfer reaction 
to $^{10}\text{Be}(0^+)$ (a) and $^{10}\text{Be}(2^+)$ (b).
The calculations for 35.3 MeV/nucleon are compared
with the experimental data from GANIL \cite{Win01}.}
\end{figure}

GANIL data for a beam of 35.3 MeV/nucleon \cite{Win01} are compared 
with our results in Fig. \ref{fig:be11pdbe10_2}.
In this case, for the calculations,
the p+\textsuperscript{11}Be potential was obtained from
the parameterization of Watson, Sigh and Segel \cite{Wat69}.
For consistency, the same parameterization is used for the construction
of the adiabatic d+\textsuperscript{10}Be potential.

Both models obtain reasonable overall agreement with the experimental data.
The Nilsson model is slightly closer to the data
when the transfer is to the ground state of \textsuperscript{10}Be (Fig.~\ref{fig:be11pdbe10_2}(a)).
However, Fig. \ref{fig:be11pdbe10_2}(b) shows how the PAMD model is still the closest to the data
for the transfer to the $2^+$ state of \textsuperscript{10}Be.

In Ref.~\cite{Jiang_2018} and~\cite{Win01},
some values of spectroscopy factors (SF) for the 
$\langle ^{10}\text{Be}|^{11}\text{Be}\rangle$ overlap were obtained from
the comparison of the measured data with ADWA calculations.
As the weights of the $\alpha$ components are considered
an approximation of these spectroscopic factors,
in Table \ref{tab:Be11_SF}, the experimental SF are compared
with the weights obtained through the Nilsson and PAMD models.
In general, the weights of both models are approximately compatible
with the ranges of experimental values for SF.
Note that there is a significant difference between the Nilsson and PAMD models
in case of $\langle^{10}\text{Be}(2^+)|^{11}\text{Be}(1/2^+_{gs})\rangle$ overlap,
due to the aforementioned difference in the component $|d_{5/2}\otimes2^+\rangle$.
This is the main reason for the discrepancy between the models for
the $^{11}\text{Be}(p,d)^{10}\text{Be}(2^+)$ differential cross section.

\begin{table}
\caption{\label{tab:Be11_SF}%
Spectroscopic factors for the $^{11}\text{Be}(p,d)^{10}\text{Be}$
reaction to the $0^+$ ground state and $2^+$ excited state in $^{10}\text{Be}$.
The values resulting from our theoretical models
are compared with the ranges of values
from the analysis of the experimental data.}
\begin{ruledtabular}
\begin{tabular}{ccccc}
 SF &\multicolumn{2}{c}{Experimental Data}&
\multicolumn{2}{c}{Theoretical Model}\\
  $|^{11}\text{Be}(1/2^+_{gs})\rangle$
  & RCNP \cite{Jiang_2018}& GANIL \cite{Win01}&Nilsson&PAMD\\
\hline
$\langle^{10}\text{Be}(0^+)|$& $0.82\pm0.15$ & 0.66-0.80 & 0.78 & 0.85\\
$\langle^{10}\text{Be}(2^+)|$& $0.26\pm0.09$ & 0.13-0.38& 0.21 & 0.15\\
\end{tabular}
\end{ruledtabular}
\end{table}

\section{\label{sec:level1}Summary and Conclusions}

A deformed two-body approach based on the Nilsson model
has been applied to the study of exotic nuclei, 
paying special attention to one-neutron halo nuclei.
This model considers a neutron moving in a 
deformed potential generated by the core.
This interaction consists of central Woods-Saxon potential,
a non-central term that assumes the permanent 
axial quadrupole deformation of the former
and a spin-orbit term.
The full Hamiltonian of the system also includes
a collective rotational term to account for core excitation.
The results of this model are compared with those of the PAMD model \cite{Lay14}.
In both cases the energies and  wave functions are obtained
by diagonalizing the Hamiltonian in
the transformed harmonic oscillator (THO) basis.
Using the adiabatic distorted wave approximation (ADWA),
the results of our two models have been applied to the study
of one-neutron transfer reactions and
their application to break up reactions is in progress.

The Nilsson and PAMD models have been applied to \textsuperscript{17}C, 
whose first excited is a one-neutron halo candidate,
and to the well-known halo nucleus \textsuperscript{11}Be.
Using a deformation parameter $\beta=0.34$,
the Nilsson model gives a good description
of the bound states in \textsuperscript{17}C,
better than the semi-microscopic PAMD model.
From the analysis of the results,
the relevance of the $4^+$ core state stands out,
a state that is not included in the PAMD model.
Furthermore, both models present a large spatial extension 
in the wave function of the first excited state,
supporting the halo nature of this state.
On the other hand, the PAMD model better reproduces
the experimental spectrum of \textsuperscript{11}Be.
However, the Nilsson model obtains similar results
by applying a deformation of $\beta=0.67$.
In both models, a parity-dependent strength is needed,
but the difference between parities is less than 6\%.

The structure models are tested by studying transfer reactions 
$^{16}\text{C}(d,p)^{17}\text{C}$ and $^{11}\text{Be}(p,d)^{10}\text{Be}$.
For the former, calculations were performed for an energy of 17.2~MeV/nucleon,
and the results were compared with the experimental data from GANIL \cite{Pereira}.
Good agreement is found with the data for the differential cross section for to the  \textsuperscript{17}C first excited state ($1/2^+_1$). 
 As discussed in \cite{Pereira}, this state is a candidate for being a halo state. Both models predict a large extension for this state and provide a good reproduction of the experimental data, thus adding more evidence to the halo nature of this excited state.
In case of the second excited state $5/2^+_1$,
our calculations are clearly below the experimental data.
This suggests that the weight of the component
$|d_{5/2}\otimes0^+\rangle$ of the $5/2^+_1$ state, 
should probably be higher than the values obtained with the two models presented here.
In general, although the results of both models are very similar,
we can conclude that the Nilsson model are closer to the data.
This work shows the transfer to the bound states of \textsuperscript{17}C,
but the transfer to the continuum states is also being studied and the results will be presented in a subsequent publication.

The $^{11}\text{Be}(p,d)^{10}\text{Be}$ transfer reaction
has been studied at two different energies,
$26.9\text{ MeV/nucleon}$ and $35.3 \text{ MeV/nucleon}$.
In the first case, the results of the reaction calculations
using the two different structure models
are compared with the RCNP data \cite{Jiang_2018} and,
in the other case, the calculations are compared 
with the experimental data from GANIL \cite{Win01}.
In both cases, for the transfer to the ground state of \textsuperscript{10}Be,
a reasonable agreement is obtained with the two models.
For the transfer to the first excited state of \textsuperscript{10}Be,
the PAMD model is closer to the data.
This is mainly due to the fact that the weight of the 
$|d_{5/2}\otimes2^+\rangle$ component is  smaller in this model.

Since the PAMD model provides better results both in the transfer cross section
and for the predicted resonant energies,
this model seems to be more adequate to describe the structure of \textsuperscript{11}Be.
However, for \textsuperscript{17}C, 
the Nilsson model gives a more accurate description of both the spectrum and the transfer reaction studied, making it a better candidate for modeling the structure of \textsuperscript{17}C.
This may be due to the inclusion of the $4^+$ core state by the Nilsson model, which was absent in the PAMD model.
The importance of the including this \textsuperscript{16}C state has been suggested by other models \cite{AMOS12}.
It can also be an indication of how, depending on the nucleus,
the strong- or weak-coupling approach between the valence nucleon and core is more appropriate.
In any case, both deformed two-body models reasonably described
the structure of \textsuperscript{17}C and \textsuperscript{11}Be.
Furthermore, because it uses the same THO formalism,
they can be just as easily included in reactions calculations.

A pending task for these two models is a more correct
application of the Pauli principle. 
Here we have removed those final bound eigenstates that we consider occupied
by comparing with the spherical and Nilsson limits.
The Nilsson model is more convenient in this regard
because it allows single-particle Nilsson states to be removed
or partially blocked in a more sophisticated way. 

The Nilsson-inspired model presented here and the PAMD model have been applied for the first time
to transfer reactions with \textsuperscript{17}C and \textsuperscript{11}Be.
Application to \textsuperscript{19}C is in progress and 
extension to other weakly bound nuclei is planned.
As both models also provide radial wavefunctions,
the halo nature of the nuclei will be consistently considered in the analysis of the reaction.
In addition, it is intended to incorporate microscopic information into the model,
in a way similar to what is done in the PAMD model or even beyond.

\begin{acknowledgments}

The present research is funded from grant PID2020-114687GB-I00 by MCIN/AEI/10.13039/501100011033, the project PAIDI 2020 with Ref. P20\_01247 by the Consejer\'{\i}a de Econom\'{\i}a, Conocimiento, Empresas y Universidad, Junta de Andaluc\'{\i}a (Spain), and by ERDF A way of making Europe. P.P. acknowledges PhD grants from the
Ministerio de Universidades and the
Consejer\'ia de Transformaci\'on Econ\'omica, Industria, Conocimiento y Universidades, Junta de Andaluc\'ia

\end{acknowledgments}

\bibliography{nilsson,c19p}

\end{document}